\documentclass[twocolumn,showpacs,preprintnumbers,amsmath,amssymb]{revtex4-1}

\usepackage{bbm}
\usepackage{subfigure}
\usepackage{amsmath}
\usepackage{epsfig,psfrag}
\usepackage{pstricks}
\usepackage{dcolumn}
\usepackage{bm}
\usepackage{graphicx}
\usepackage{color}
\usepackage{version}
\usepackage[normalem]{ulem}

\newcommand{\av}[1]{\left<#1\right>}
\newcommand{\CKeldysh}{\mathcal C}

\begin{document}
\title{ Tunneling into quantum wires: regularization of the tunneling Hamiltonian and consistency between free and bosonized fermions}
\author{Michele Filippone}
\author{Piet Brouwer}
\affiliation{Dahlem Center for Complex Quantum Systems and Institut f\"ur Theoretische Physik, Freie Universit\"at Berlin, Arnimallee 14, 14195 Berlin, Germany}

\begin{abstract}
{
Tunneling between a point contact and a one-dimensional wire is usually described with the help of a tunneling Hamiltonian that contains a delta function in position space. Whereas the leading order contribution to the tunneling current is independent of the way this delta function is regularized, higher-order corrections with respect to the tunneling amplitude are known to depend on the regularization. Instead of regularizing the delta function in the tunneling Hamiltonian, one may also obtain a finite tunneling current by invoking the ultraviolet cut-offs in a field-theoretic description of the electrons in the one-dimensional conductor, a procedure that is often used in the literature. For the latter case, we show that standard ultraviolet cut-offs lead to different results for the tunneling current in fermionic and bosonized formulations of the theory, when going beyond leading order in the tunneling amplitude. We show how to recover the standard fermionic result using the formalism of functional bosonization and revisit the tunneling current to leading order in the interacting case. }
\end{abstract}


\maketitle

\section{Introduction}

When electrons are confined to a single spatial dimension, the screening of interactions becomes much less effective, and the description in terms of a ``Fermi liquid'' of effectively non-interacting particles breaks down. Instead, at low energies, interactions are responsible for the emergence of bosonic collective excitations, which are described by the Luttinger liquid theory \cite{tomonaga50,luttinger63,mattis65,haldane81,haldane81prl,giamarchi04}. The description of this many-electron system in terms of these collective modes is called \textit{bosonization} \cite{haldane81,haldane81prl}. Technically, bosonization is understood as an operator identity between fermionic and bosonic operators in one dimension \cite{vondelft98,schonhammer97,schonhammer04}. 

The confinement of electrons to one dimension has been achieved in a variety of solid state devices. Examples are carbon nanotubes \cite{bockrath99}, cleaved-edge overgrowth wires in semiconductor heterostructures \cite{tarucha95,yacoby96,auslaender02,auslaender05}, metallic chains in nanowires \cite{slot04,venkataram06}, polymer nanofibers \cite{aleshin04}, or the edge states of the Quantum Hall insulator \cite{wen90,chang03}. All of these systems provide the possibility to investigate the properties of this interaction-dominated state of matter for which the Fermi liquid theory \cite{abrikosov88,nozieres99} does not apply. 

A particular problem of interest is the tunneling of electrons into one-dimensional interacting wires. At low temperatures, the tunnel current has a power-law dependence on the applied bias, which depends on the strength of interactions in the one-dimensional wire \cite{kane97,yao99,eggert00}. Tunneling experiments allowed the observation of fractional charge carriers in the edge states of fractional Quantum Hall states \cite{kane94,fendley95,chamon95} in noise measurements \cite{saminadayar97,depicciotto97}. Moreover, recent experimental \cite{chen09,altimiras10} and theoretical \cite{gutman10} studies have shown how tunnel junctions can also probe the out-of-equilibrium distribution of electrons in one dimensional strongly correlated systems. 

Tunneling through a point-like contact (in contrast to the momentum-conserving tunneling in cleaved-edge-overgrowth wires \cite{yacoby96,auslaender02,auslaender05}) is usually described by a tunneling Hamiltonian
\begin{align}\label{eq:gamma}
  \mathcal H_\gamma &=\gamma\psi^\dagger(0)c(0)+\gamma^*c^\dagger(0)\psi(0)\,,
\end{align}
where $\gamma$ is the tunneling amplitude, $\psi(x)$ and $\psi^{\dagger}(x)$ are (fermionic) annihilation and creation operators for the interacting one-dimensional wire, and $c(x)$ and $c^{\dagger}(x)$ are (fermionic) annihilation and creation operators for the reservoir, which is taken to be effectively non-interacting. The tunneling point contact is at position $x=0$ and for simplicity we consider a spinless system. To lowest nontrivial order in the tunneling amplitude, the zero-temperature tunneling current $I$ reflects the suppression by interactions of the single particle density of states in the wire \cite{giamarchi04,voit95,schonhammer97,schonhammer04,fisher97}
\begin{equation}
  I = \frac{|\gamma^2| e^2}{2 \pi u v \hbar^3} V \left( \frac{eV}{\Lambda} \right)^{\frac12\left(K+1/K\right)-1},
  \label{eq:2}
\end{equation}
where $V$ is the applied bias, $u$ and $v$ the Fermi velocities in the wire and the reservoir, respectively, $K$ is the ``Luttinger parameter'', which equals one in the absence of interactions and satisfies $K < 1$ ($K > 1$) for repulsive (attractive) interactions, and $\Lambda$ is an energy scale set by the interactions.

For certain applications it is important to go beyond the lowest order in the tunneling amplitude $\gamma$. An example is the calculation of tunneling currents beyond linear response \cite{schoeller95}, but the inclusion of higher orders in $\gamma$ may also be relevant for calculations of the shot noise or for Andreev processes \cite{jong94}, for which the tunneling effectively occurs for pairs of electrons. For such higher-order processes the tunneling Hamiltonian (\ref{eq:gamma}) is no longer a well-defined starting point and a regularization with respect to the position $x$ of the tunneling point contact is needed. In general the higher-order contributions to the tunneling current depend on this regularization. Although this is known to the experts in the field, what surprised us and prompted us to write the present article, was our finding that a calculation without a regularization of the tunneling Hamiltonian nevertheless leads to a finite and convergent answer, albeit in such a way that the result of a fermionic calculation differs from that obtained by standard application of the bosonization identities. 

Since the dependence on the regularization procedure and the differences between fermionic and bosonized approaches already exist on the non-interacting level (Luttinger parameter $K=1$), most of our article will consider this special case. We also consider an alternative bosonization procedure, known as ``functional bosonization'' \cite{yurkevich02,lerner05}, and show that it resolves the inconsistency between fermionic and bosonic descriptions in a quite elegant and direct way --- although the fundamental dependence on the choice of regularization for the tunneling Hamiltonian (\ref{eq:gamma}) continues to exist.

The necessity to regularize tunneling Hamiltonians applies to all studies addressing the strong tunneling limit for electrons in one-dimensional interacting systems. This includes impurity problems out of equilibrium, for which bosonization, followed by a ``refermionization'', allowed major theoretical breakthroughs. A notable example is the Kondo problem at the Toulouse limit \cite{emery92,schiller95}. In this limit, a fine tuning of hopping parameters in the Kondo problem allows to map it on a free fermion problem. Recently, Shah and Bolech showed that the naive utilization of the bosonization/refermionization identities leads to qualitative deviations from the correct result in the strong tunneling limit \cite{shah15,bolech15}. Whereas these authors advocate an ad-hoc modification of the refermionization schemes to amend the issue, our results suggest that the inconsistency may already appear at the level of the bosonization procedure itself.

\begin{figure}
\begin{center}
\includegraphics[width=.4\textwidth]{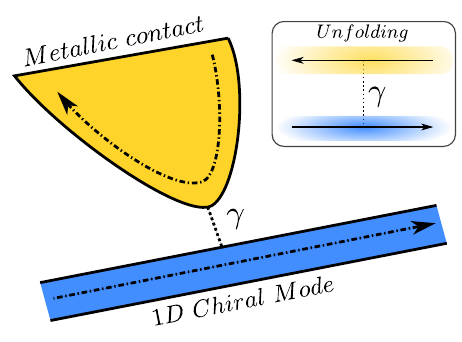}
\caption{Tunneling contact between a metallic contact and an interacting one-dimensional wire. In most of our discussion we take the wire to have a single chiral ({\em i.e.}, unidirectional) mode. The inset shows a schematic representation of the unfolding procedure that allows the electrons in the metallic contact to be described as a one-dimensional chiral mode. }\label{fig:tip}
\end{center}
\end{figure}

We start in Sec.\ \ref{sec:model} with a formulation of the model Hamiltonian that describes tunneling between a Fermi-liquid reservoir and a Luttinger liquid. We solve the out-of-equilibrium problem in the non-interacting case by relying on standard scattering theory \cite{landauer81,*buttiker85} and illustrate how 
different regularizations of the tunneling term imply different qualitative behaviors for the conductance.  Upon expanding in the tunneling amplitude $\gamma$, the differences between different regularizations do not appear to leading (second) order in the tunneling amplitude $\gamma$, but only to next-to-leading (fourth) order correction to the conductance. In Sec.\ \ref{sec:standbos}, we consider a calculation of the tunneling current for an unregularized tunneling Hamiltonian \cite{kane94,fendley95,chamon95,safi01,*guyon02,*guyon02a,crepieux03,pugnetti09}, but with a short-distance cut-off in the fermionic or bosonic propagators, and show that the fermionic and bosonic versions of the non-interacting theory lead to different but convergent results for the tunneling currents. We also show that such a difference does not occur if the tunneling Hamiltonian is regularized. In Sec.\ \ref{sec:funcbos}, we show that functional bosonization with Luttinger Liquid parameter $K=1$ is consistent with the free fermion result to all order in the tunneling amplitude. We also discuss how functional bosonization allows an intuitive distinction between interaction cutoffs, necessary to regularize the bosonized theory, and short-distance cutoffs, necessary to regularize free fermions with linear dispersion. Finally, in Sec.\ \ref{sec:int}, we provide an illustration about how this distinction affects the current in the presence of interactions in the wire. 




\section{Model}\label{sec:model}

\begin{figure}
\begin{center}

\includegraphics[width=.38\textwidth]{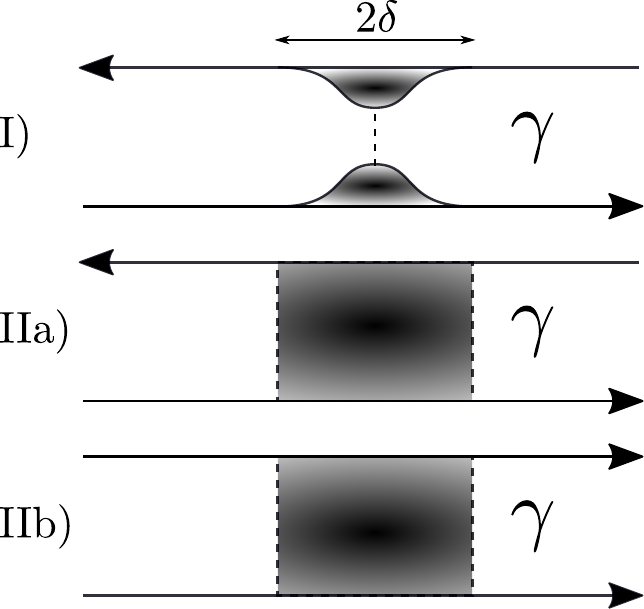}
\caption{Sketch of the regularization schemes Eqs. \eqref{eq:ch1} and \eqref{eq:ch2} for the tunneling coupling Eq. \eqref{eq:gamma}. Choice II is sensitive to the respective chirality of electrons in the wire and in the contact and we make a distinction between Choice IIa and IIb.}\label{fig:regs}
\end{center}
\end{figure}

In Fig. \ref{fig:tip}, we picture a typical device considered for the study of the tunneling of  electrons in 1D wires. It is composed of a metallic contact, such as the tip of a scanning probe or an integer quantum Hall edge state, tunnel-coupled to a spinless one-dimensional wire. By a standard unfolding procedure, electrons in the contact, which are taken to be noninteracting, can be always mapped onto a chiral one-dimensional system \cite{fabrizio95}. For electrons in the wire we take a Luttinger-Liquid description. To keep the discussion at the simplest possible level, we focus most of our discussion on the situation in which the wire has a single chiral ({\em i.e.}, unidirectional) spinless mode.

The Hamiltonian for this system consists of three terms, 
\begin{align}\label{eq:hamsys}
\mathcal H=\mathcal H_{\rm C}+\mathcal H_{\rm W }+\mathcal H_{\rm \gamma}\,,
\end{align} 
where $H_{\rm C}$ and $H_{\rm W}$ describe the electrons in the contact and the wire, respectively. Without interactions in the one-dimensional wire they read 
\begin{align}
\mathcal H_{\rm C} &= r\, \int_{-\infty}^\infty dx\, c^\dagger(x)(-i\hbar v\partial_x)c(x)\,, \label{eq:contact}\\ 
\label{eq:wire}\mathcal H_{\rm W }&= \int_{-\infty}^\infty dx\, \psi^\dagger(x)(-i\hbar u\partial_x)\psi(x)\,,
\end{align}
where the operators $c(x)$ and $\psi(x)$ describe electrons in the contact and the wire, respectively. The term $H_{\gamma}$ describes tunneling between the contact and the wire; it is given in Eq.\ (\ref{eq:gamma}). The prefactor  $r=\pm$ in Eq. \eqref{eq:contact} sets the propagation direction of the electrons in the contact with respect to the wire, a detail of some importance for certain regularization procedures for the tunneling Hamiltonian $H_{\gamma}$, to be discussed below.

\begin{figure}
\begin{center}
\includegraphics[width=.5\textwidth]{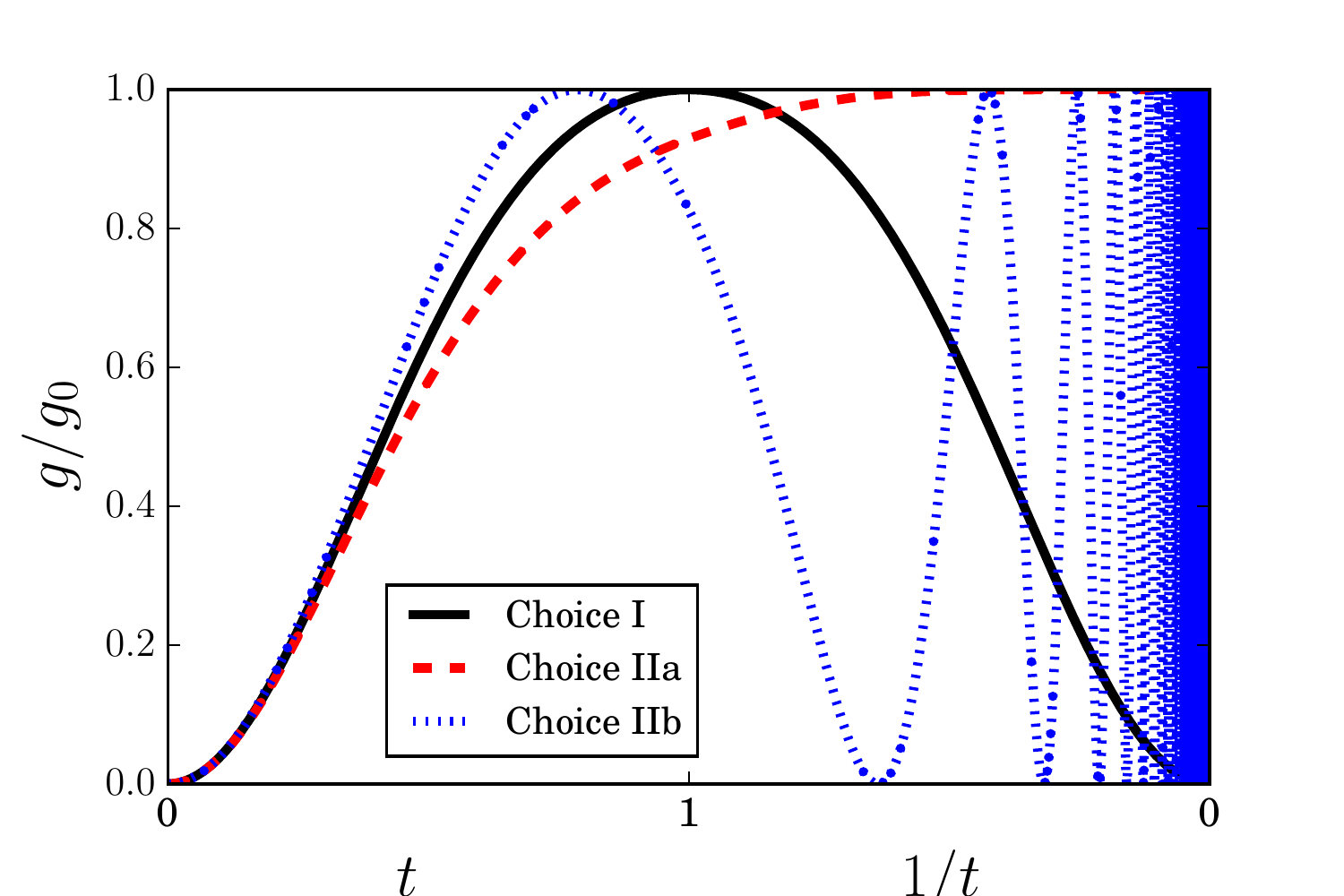}
\caption{Conductances in Eq. \eqref{eq:conductances} as a function of the tunneling parameter $t=\gamma/2\hbar\sqrt{uv}$. Different choices of the tunneling regularization are responsible for different qualitative behaviors in the limit $t\rightarrow \infty $, even if they all have the same behavior for $t\rightarrow 0$. }\label{fig:conductances}
\end{center}
\end{figure}

As long as interactions are neglected in the wire, the Hamiltonian \eqref{eq:hamsys} is quadratic in the fermion creation and annihilation operators and the stationary current induced by a voltage bias $V$ between the contact and the wire can be calculated exactly from standard scattering theory. To second order in $\gamma$ the result is unambiguous; when higher orders are included, a regularization of the tunneling Hamiltonian must be specified. We focus on two possible choices, 
\begin{align}
\mbox{Choice I:}&& \begin{split}
\label{eq:ch1}\psi(0) &\rightarrow\int dx f(x)\psi(x)\ \\ 	
  \mbox{and}\ c(0) &\rightarrow \int dx f(x) c(x),
\end{split}\\
\label{eq:ch2}\mbox{Choice II:}&& \psi^\dagger(0)c(0) & \rightarrow\int dx f(x)\psi^\dagger(x)c(x)\,,
\end{align}
with $f(x) = (1/2 \delta) \Theta(\delta-|x|)$, $\Theta(x)$ being the Heaviside step function and $\delta$ the regularization scale. For the second choice it matters whether the chiral modes in the wire and the contact have the same propagation direction, and we refer to the two cases as ``IIa'' and ``IIb'', corresponding to $r=-1$ and $r=1$ in Eq.\ (\ref{eq:contact}), respectively, see Fig.\ \ref{fig:regs}. Without interactions, the relation between current and applied bias $V$ is always linear. However, the conductance $G = I/V$ depends on the choice of the regularization. The resulting expressions for the conductance contain the tunneling amplitude $\gamma$ in the combination
\begin{equation}
  t= \frac{\gamma}{2\hbar\sqrt{uv}}, \label{eq:tdef}
\end{equation}
and they read (the explicit derivation can be found in Appendix A): 
\begin{subequations}\label{eq:conductances}
\begin{align}
\label{eq:I1} G^{({\rm I})} &= \frac{e^2}{h}\frac{4t^2}{(1+t^2)^2}\,,\\
\label{eq:I2} G^{({\rm IIa})} &=\frac{e^2}{h}\tanh^2(2t)\,,\\
\label{eq:I3} G^{({\rm IIb})} &=\frac{e^2}{h}\sin^2(2t)\,.
\end{align}
\end{subequations} 
The three expressions are plotted in Fig. \ref{fig:conductances} as a function of the tunneling parameter $t$. Although the three expressions coincide up to order $t^2$, 
\begin{align}\label{eq:ilot}
 G &= \frac{4 e^2 t^2}{h} + {\cal O}(t^4),
\end{align}
the three regularization schemes differ rather strongly in the limit $t\rightarrow \infty$ \cite{shah15,bolech15}.

In the case of Choice I, the suppression of the conductance in the $t\rightarrow \infty$ limit can be explained by the formation of bonding and anti-bonding states at the junction between electrons in the contact and in the wire. This leads to a local suppression of the density of states, reducing the conductance. In the case of Choice II, the situation is quite different. As already mentioned, Choice II is sensitive to the  respective chirality of the contact and the wire. For opposite chiralities (IIa), in the high hybridization limit, $H_{\gamma}$ opens up a gap in the contact and the wire, leading to complete backscattering and, hence, complete transmission into the wire, in the limit $t \to \infty$. For equal chiralities (IIb), on the other hand, electrons oscillate coherently between contact and wire, and the total transmission depends sensitively on the length of the contact.


\section{Comparison with standard bosonization}\label{sec:standbos}

\subsection{Perturbation theory}

The three explicit examples discussed in the previous Section clearly demonstrate the importance of the regularization scheme for the calculation of the conductance to arbitrary order in the tunneling strength $t$. Nevertheless, it is worth asking to which extent it is possible to avoid the specification of the regularization of the tunneling term \eqref{eq:gamma}. The practice of not specifying the regularization is widespread in the literature, in particular for wires with interacting electrons \cite{kane94,fendley95,chamon95,safi01,*guyon02,*guyon02a,crepieux03,pugnetti09}, for which the exact solution of Sec.\ \ref{sec:model} is not available.

What makes it possible to avoid specifying the regularization for the tunneling Hamiltonian is that standard approaches in terms of propagators for the electrons in the wire and the contact involve an {\em additional} regularization, which seemingly appears to remove the necessity to regularize the tunneling term. This short-distance cut-off appears in both the fermionic and the bosonic formulation of the theory. We will now calculate the conductance to fourth order in the tunneling amplitude $\gamma$ for both formulations separately and show that they lead to different results. 

By a standard gauge transformation, it is possible to absorb the difference $e V$ of the chemical potentials in the contact and the wire in the tunneling term Eq. \eqref{eq:gamma}, by replacing 
\begin{equation}
  \label{eq:gammat}
  \gamma\rightarrow \gamma(t) =\gamma e^{i\Omega t},\ \mbox{with}\
  \Omega = e V/\hbar.
\end{equation}
The coupling term of Eq. \eqref{eq:gamma} then acts as a time dependent perturbation to the system. The current operator reads 
\begin{align}
\hat I(t)=i\frac e\hbar\Big[\gamma(t)\psi^\dagger(0,t)c(0,t)-\gamma^*(t)c^\dagger(0,t)\psi(0,t)\Big]
\end{align}
and the tunneling current is readily written in terms of a correlation function between fermions in the contact and in the wire
\begin{align}\label{eq:currgreen}
  I = \frac{2e}\hbar\mbox{Re}\left[\gamma(t)\Lambda^<(t,t;0,0)\right]\,,
\end{align}
in which 
\begin{equation}
  \Lambda^{\CKeldysh}(t,t';x,x)=-i\av{\mathcal T_{\CKeldysh} c(x,t)\psi^\dagger(x',t')}.
  \label{eq:Lambdadef}
\end{equation}
Here we have introduced the ordering $\mathcal T_{\CKeldysh}$ along the Keldysh contour $\mathcal C_{\rm K}$ \cite{kamenev11}. Times $t^\pm$ evolve on an upper/lower branch $\mathcal C^{\pm}$ of this contour, and we adopt the convention that the upper/lower branch runs forward/backward in time. The function $\Lambda^{\CKeldysh}$ equals the time-ordered Green function $\Lambda^{\rm T}$ if both $t$ and $t'\in\mathcal C^+$, it equals the anti-time ordered Green function $\Lambda^{\tilde {\rm T}}$ if both $t$ and $t' \in \mathcal C^{-}$, it equals the greater Green function $\Lambda^>$ if $t\in \mathcal C^-$ and $t'\in \mathcal C^+$, and the lesser Green function $\Lambda^<$ if $t\in \mathcal C^+$ and $t'\in \mathcal C^-$. 

Using perturbation theory in $H_{\gamma}$, the correlation function $\Lambda^{\CKeldysh }$ and, hence, the current $I$ can be expressed in terms of the correlation functions of the fermions in the wire and the contact, which have to be evaluated without the tunneling Hamiltonian $H_{\gamma}$. Explicitly, up to fourth order in the tunneling amplitude, one has
\begin{equation}
  I = I^{(2)} + I^{(4)},
\end{equation}
with
\begin{widetext}
\begin{align}\label{eq:ilo}
  I^{(2)} =& \, \frac{2 e \gamma^2}{\hbar^2} 
  \mbox{Re} \left[
  \int_{\mathcal C_{\rm K}} dt_1 e^{i \Omega(t-t_1)}
  \av{-i \mathcal T_{\CKeldysh} c(t^+) c^{\dagger}(t_1)}
  \av{-i \mathcal T_{\CKeldysh} \psi(t_1) \psi^{\dagger}(t^-)}
  \right], \\
  I^{(4)} = & \, \mbox{} -\frac{e \gamma^4}{\hbar^4}
  \mbox{Re}\left[\int_{\mathcal C_{\rm K}} dt_1dt_2dt_3\,e^{i\Omega(t+t_1-t_2-t_3)}
  \av{- \mathcal T_{\CKeldysh} c(t^+)c(t_1)c^\dagger(t_2)c^\dagger(t_3)}
  \av{- \mathcal T_{\CKeldysh}\psi(t_3)\psi(t_2)\psi^\dagger(t_1)\psi^\dagger(t^-)}
  \right],
  \label{eq:ilo4}
\end{align}
\end{widetext}
where $t^{\pm}$ is the point on ${\cal C}^{\pm}$ corresponding to the real time $t$. To keep the expressions compact, we have suppressed the spatial argument $x=0$ for the fields $\psi(x,t)$ and $c(x,t)$. The pair correlation functions in Eq.\ (\ref{eq:ilo}) are nothing but the contour-ordered Green functions $G^{\CKeldysh}(x-x';t-t')$ for electrons in the wire,
\begin{equation}\label{eq:greenfree}
\begin{aligned}
  G^{\CKeldysh}(x&-x',t-t') \\
  &= -i\av{\mathcal T_{\CKeldysh} \psi(x,t)\psi^\dagger(x',t')}\\ &= \frac1{2 \pi [x-x'-u(t-t')+i \alpha\, s_{\CKeldysh}(t-t')]}\,,
\end{aligned}
\end{equation}  
where $s_{\CKeldysh}(t-t')$ is the ``contour-ordered sign'', 
\begin{equation}
s_{\CKeldysh}(t-t')=\left\{\begin{array}{ll}
\mbox{sign}(t)&~~~t,t'\in\mathcal C^+,\\
+1&~~~t\in\mathcal C^-,t'\in\mathcal C^+,\\
-1&~~~t\in\mathcal C^+,t'\in\mathcal C^-,\\
-\mbox{sign}(t)&~~~t,t'\in\mathcal C^-,
\end{array}
\right.
\end{equation}
and $\alpha$ is a short-distance cut-off that must be sent to zero at the end of the calculation. Physically, $\alpha$ represents the finite band width for the fermionic fields, which should plays no role for phenomena taking place in the immediate vicinity of the Fermi level. The expression for the Green function $C^{\CKeldysh}(x-x';t-t') = -i\av{\mathcal T_{\CKeldysh} c(x,t)c^\dagger(x',t')}$ for the electrons in the contact is identical, up to the replacement $x,x' \to r x, r x'$.

\subsection{Fermionic approach}

In the fermionic formulation, higher-order correlation functions, such as those that appear in Eq.\ (\ref{eq:ilo4}) can be expressed in the function (\ref{eq:greenfree}) using Wick's theorem,
\begin{align}
 & \av{- {\mathcal T_{\CKeldysh}} \psi(t_1)\psi(t_2)\psi^\dagger(t_3)\psi^\dagger(t_4)} \nonumber
  \\  &~~~~=  G^{\CKeldysh}(0,t_1-t_4) G^{\CKeldysh}(0,t_2-t_3) \nonumber \\ &  ~~~~~~~~~ 
  -G^{\CKeldysh}(0,t_1-t_3) G^{\CKeldysh}(0,t_2-t_4). \label{eq:4pfermion}
\end{align}
Because of the presence of the short-distance cut-off $\alpha$, the current \eqref{eq:currgreen} can be calculated to any order in $\gamma$. Since the tunneling term is not regularized, the chirality $r$ of the fermions in the contact does not play a role in the calculation. After taking the limit $\alpha \downarrow 0$ at the end of the calculation, the result coincides with Eq. \eqref{eq:I1} ({\em i.e.,} our ``Choice I''), without the apparent need of a regularization of the tunneling coupling Eq.\ \eqref{eq:gamma} \cite{kakshvili08}. In particular, for the second-order and fourth-order contributions to the tunneling current we find
\begin{align}
  I^{(2)} & = \frac{4 t^2 e^2 V}{h}, \label{eq:i2fermion} \\
  I^{(4)} & = - \frac{8 t^4 e^2 V}{h}. \label{eq:i4fermion}
\end{align}

\subsection{Bosonization approach}

The starting point for a calculation using the bosonization formalism is the identity \cite{haldane81,haldane81prl,giamarchi04,vondelft98}
\begin{equation}\label{eq:bos}
  \psi(x)=\frac {F}{\sqrt{2\pi a}}e^{-i\phi(x)}\,,
\end{equation}
which expresses the (chiral) fermion operator $\psi(x)$ in terms of a (chiral) bosonic field $\phi$ and a ``Klein factor'' $F$. Strictly speaking, Eq.\ (\ref{eq:bos}) applies to an infinite system size only \cite{vondelft98}. The short-distance cutoff $a$ ensures convergence of correlation functions of the bosonic fields, but is a priori not necessarily identical to the cutoff $\alpha$ appearing in the free fermion correlation functions \eqref{eq:greenfree} \cite{zarand98,*zarand00}. The boson fields are subject to the Hamiltonian
\begin{equation}\label{eq:hamphi}
  H_{\phi} = \frac{\hbar u}{4\pi}\int dx \big(\partial_x \phi\big)^2.
\end{equation}
Right moving fields obey the Kac-Moody relation $[\partial_x\phi(x),\phi(x')]=2\pi i \delta(x-x')$, which is fulfilled by 
\begin{equation}
\phi(x)=-i\sum_{q>0}\sqrt{\frac{2\pi}{Lq}}\left[b_qe^{iqx}-b_q^\dagger e^{-iqx}\right]\,,
\end{equation}
in which $[b_q,b^\dagger_q]=1$, allowing to diagonalize Eq.\ \eqref{eq:hamphi} and to derive the two-point correlation function 
\begin{equation}
\begin{aligned}
  \frac12&\av{\mathcal T_{\CKeldysh} \big[\phi(x,t)- \phi(x',t')\big]^2} \\& = \ln\left[1-\frac{is_{\CKeldysh}(t-t')[(x-x')-u(t-t')]}a\right]\,,
\end{aligned}
\end{equation}
in which we introduced the bosonic cutoff `$a$' to regularize sums over momenta. 
Combined with the correlation function of the Klein factors,
\begin{equation}
  \av{ \mathcal T_{\CKeldysh} F(t) F^\dagger(t')} = s_{\CKeldysh}(t-t'),
\end{equation}
the bosonized theory precisely reproduces the Green function (\ref{eq:greenfree}), with the substitution $\alpha \to a$ \cite{giamarchi04,vondelft98,fradkin91} --- a possible a posteriori reason to equate the two short-distance cut-offs. It then follows directly, that the bosonized and fermionic formulations lead to the same tunneling current $I$ to second order in the tunneling amplitude $t$.

For the fourth-order contribution to the tunneling current $I^{(4)}$, we need to evaluate the four-point correlation function $\av{-\mathcal T_{\CKeldysh}\psi(t_1)\psi(t_2)\psi^\dagger(t_3)\psi^\dagger(t_4)}$ in the bosonized formalism, see Eq.\ (\ref{eq:ilo4}). (We continue to use the fermionic formalism for the electrons in the contact.) One finds \cite{vondelft98}
\begin{widetext}
\begin{align}\label{eq:4p}
  \av{-\mathcal T_{\CKeldysh}\psi(t_1)\psi(t_2)\psi^\dagger(t_3)\psi^\dagger(t_4)}
  &=
  \frac{\av{-\mathcal T_{\CKeldysh} F(t_1)F(t_2)F^\dagger(t_3)F^\dagger(t_4)}}
  {(2\pi a)^2}
  \frac{f_{12}f_{34}}{f_{13}f_{14}f_{23}f_{24}}\,,
\end{align}
where we abbreviated
\begin{equation}\label{eq:f}
f_{12} =1+i\frac {u (t_1-t_2)}a\,s_{12},\ \ \mbox{and}\ \
  s_{ij} = s_{\CKeldysh}(t_i-t_j).
\end{equation}
The contour-ordered expectation value of the four Klein factors is given by Wick's theorem \cite{vondelft98,guyon02a}
\begin{equation}
\langle-\mathcal T_{\CKeldysh} F(t_1)F(t_2)F^\dagger(t_3)F^\dagger(t_4)\rangle = s_{13} s_{24} - s_{12} s_{34} - s_{14} s_{23}.
\end{equation}
It is interesting to compare Eq.\ (\ref{eq:4p}) with the four-point correlation function (\ref{eq:4pfermion}) obtained in the fermionic approach. Hereto we apply the equality
\begin{align}\label{eq:sign}
\frac{s_{12} s_{34}+s_{14}s_{23}-s_{13}s_{24}}{s_{12}s_{34}s_{13}s_{14}s_{23}s_{24}}=1\,, 
\end{align}    
and a simplified form of the Cauchy identity \cite{zinn02,mussardo09}
\begin{equation}\label{eq:cauchy}
\begin{aligned}
  \frac{-(t_1-t_2)(t_3-t_4)}{(t_1-t_3)(t_1-t_4)(t_2-t_3)(t_2-t_4)}
  =\frac1{t_1-t_3}\frac1{t_2-t_4}-\frac1{t_1-t_4}\frac1{t_2-t_3}.
\end{aligned}
\end{equation}
The four-point correlation function (\ref{eq:4p}) can then be cast into the form
\begin{equation}\label{eq:nowick}
\begin{split}
\av {- {\mathcal T_{\CKeldysh}}
  \psi(t_1)\psi(t_2)\psi^\dagger(t_3)\psi^\dagger(t_4)}= \mbox{} &
  G^{\CKeldysh}(0,t_1-t_4) G^{\CKeldysh}(0,t_2-t_3)
  \frac{\left[1-i\frac{a}{s_{12}u(t_1-t_2)}\right]\left[1-i\frac{a}{s_{34}u(t_3-t_4)}\right]}{\left[1-i\frac{a}{s_{13}u(t_1-t_3)}\right]\left[1-i\frac{a}{s_{24}u(t_2-t_4)}\right]}
  \\&
 - G^{\CKeldysh}(0,t_1-t_3) G^{\CKeldysh}(0,t_2-t_4)
  \frac{\left[1-i\frac{a}{s_{12}u(t_1-t_2)}\right]\left[1-i\frac{a}{s_{34}u(t_3-t_4)}\right]}{\left[1-i\frac{a}{s_{14}u(t_1-t_4)}\right]\left[1-i\frac{a}{s_{23}u(t_2-t_3)}\right]}\,,
  \end{split}
\end{equation}
\end{widetext}
where the short-distance cut-off in the Green function $G^{\CKeldysh}$ should be taken equal to the short-distance cut-off $a$ of the bosonized theory. Since the expressions between brackets become unity if the short-distance cut-off $a$ is sent to zero, this expression coincides with Eq. (\ref{eq:4pfermion}) obtained from the fermionic theory if the short-distance cut-offs $\alpha$ and $a$ are both sent to zero, which is possible if the four times $t_1$, $t_2$, $t_3$, and $t_4$ all have different values. This is also the requirement to recover the generalized form of Wick's theorem for interacting fields in one dimension \cite{francesco12,vondelft98}. 

\begin{figure}
\begin{center}
\includegraphics[width=0.5\textwidth]{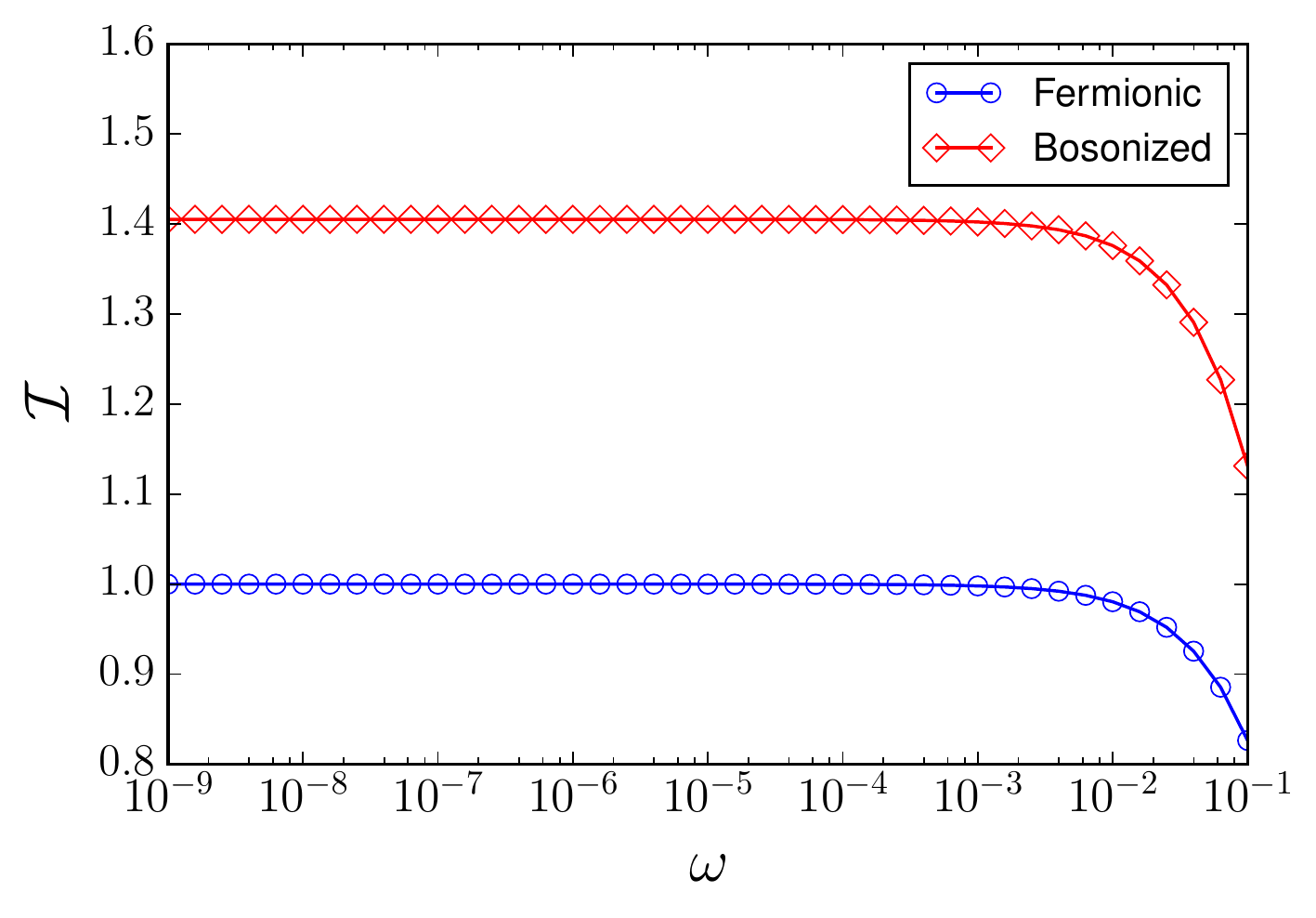}
\caption{Numerical calculation of $\mathcal I$ in Eq. \eqref{eq:nloi} as a function of $\omega = \alpha\Omega/v$ for fixed $v a/u \alpha=1$. Diamonds are calculated starting from Eq. \eqref{eq:nowickcurrent}, using the standard bosonization approach. Circles are calculated using the fermionic approach, using Wick's theorem for the four-point correlators in Eq.\ (\ref{eq:ilo4}). The estimated errors of the numerical results are less than the symbols used to represent the data points. The limit $\alpha \to 0$ clearly differs for the two approaches.}\label{fig:discrepancy}
\end{center}
\end{figure}

\begin{figure}
\begin{center}
\includegraphics[width=0.5\textwidth]{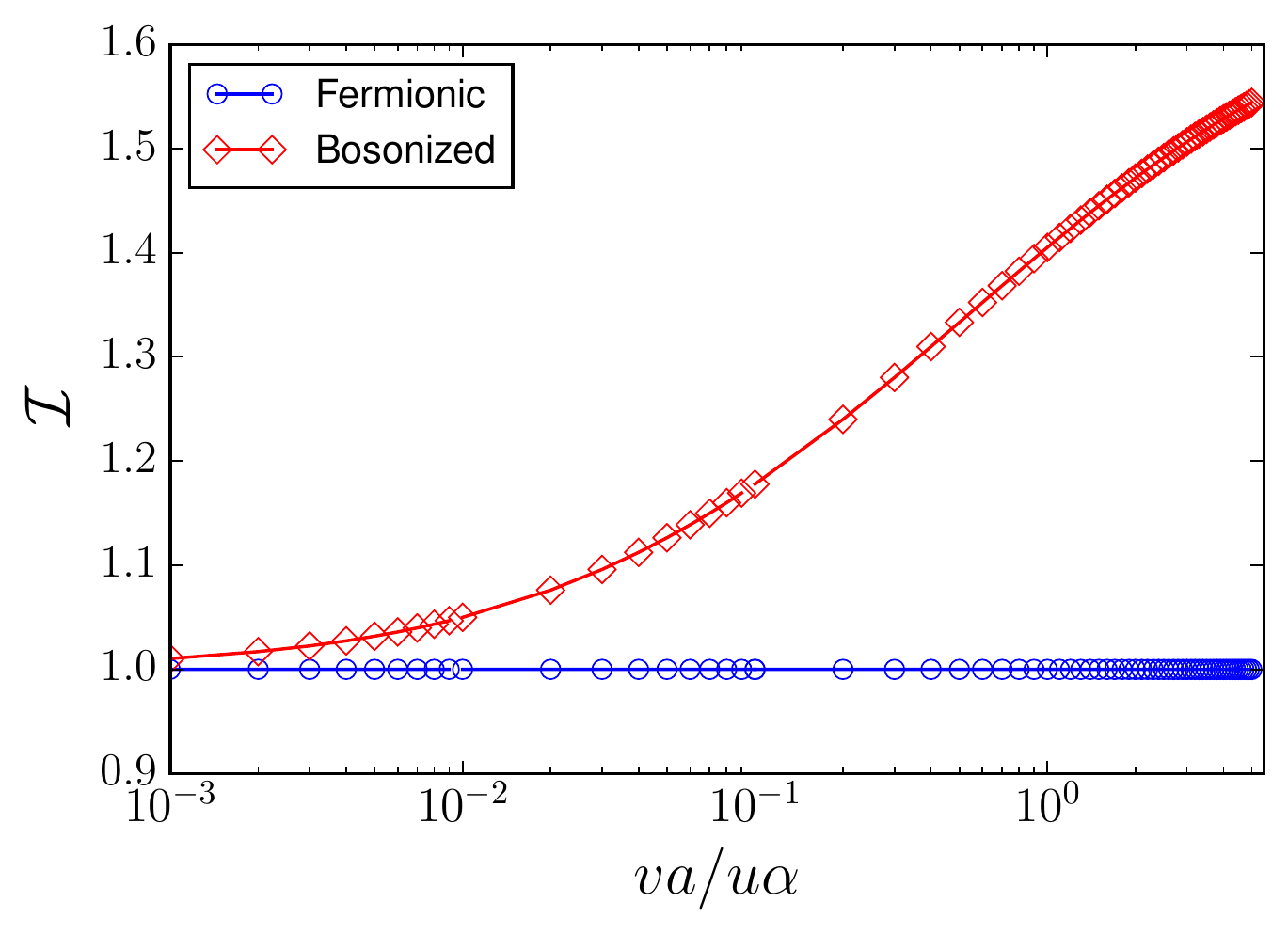}
\caption{Numerical calculation of $\mathcal I$ as a function of $v a/u \alpha$. Diamonds are for the standard bosonization approach. }\label{fig:ldep}
\end{center}
\end{figure}

The expressions (\ref{eq:4pfermion}) and (\ref{eq:nowick}) are not identical if two or more of the times $t_1$, $t_2$, $t_3$, and $t_4$ coincide, or differ less than the short-distance cut-off $a$. We now show that this difference has consequences for physical observables calculated from the correlation function. In particular, calculating the fourth-order-in-tunneling contribution to the current $I^{(4)}$ of Eq.\ (\ref{eq:ilo}) we find, with the help of Eq.\ (\ref{eq:nowick}), that
\begin{align}\label{eq:nloi}
 I^{(4)}= - \frac{8 t^4 e^2 V}{h} {\cal I},
\end{align}
with
\begin{widetext}
\begin{align}
  {\cal I} = \mbox{} & -
  \lim_{\omega \to 0} \frac{1}{2 \pi^3 \omega}\mbox{Re}\int_{-\infty}^{\infty} d\tau_1d\tau_2d\tau_3\sum_{\eta_1\eta_2\eta_3 = \pm}
\eta_1\eta_2\eta_3 e^{i\omega(\tau_1-\tau_2-\tau_3)}
\frac{s_{+2}s_{13}(s_{32}s_{1-}+s_{3-}s_{21}-s_{31}s_{2-})
}{(-i\tau_2s_{+2}+1)(i(\tau_1-\tau_3)s_{13}+1)
} \nonumber \\ & \mbox{} \times
  \frac{f'(\tau_3-\tau_2)f'(\tau_1-\tau^-)}{f'(\tau_3-\tau_1)f'(\tau_3-\tau^-)f'(\tau_2-\tau_1)f'(\tau_2-\tau^-)},
  \label{eq:nowickcurrent}
\end{align}
\end{widetext}
where $\omega=\alpha\Omega/v$, $\eta_j$ denotes the upper/lower branch of the Keldysh contour corresponding to the real time $\tau_j$, $j=1,2,3$, and
\begin{align}\label{eq:f1}
f'(t_k-t_l)&=\frac{v a}{u \alpha} +i(\tau_k-\tau_l)s_{kl}\,
\end{align}
a dimensionless version of Eq.\ \eqref{eq:f}, with $\alpha$ the short-time cut-off for the (fermionic) states in the contact.
We have not been able to carry our the remaining integration analytically, but we could perform the integral numerically using Monte Carlo sampling, which gives the result
\begin{equation}
  {\cal I} \approx 1.4,
\end{equation}
for $u = v$ and $a = \alpha$, see Fig.\ \ref{fig:discrepancy}, with quite good convergence for the limit $\omega \to 0$. For comparison, the fermionic approach gives ${\cal I} = 1$, see Eq.\ (\ref{eq:i4fermion}). Also, in the bosonization formalism the current $I^{(4)}$ depends on the ratio $u \alpha/v a$, see Fig.\ \ref{fig:ldep}, whereas there is no such dependence in the fermionic calculation based on Wick's theorem.  To this order in the tunneling amplitude, the results from bosonized and fermionic calculations coincide in the limit $u \alpha/v a \to \infty$.

The discrepancy between the fermionic and bosonized approaches, despite the mathematical equivalence between the two methods \cite{vondelft98}, suggests that the introduction of the short distance cutoff somehow corresponds to an uncontrolled regularization of the tunneling term \eqref{eq:gamma}.  In Section \ref{app:reg}, we show that both approaches give the same expression for the tunneling current to next-to-leading order in $\gamma$ if the tunneling Hamiltonian (\ref{eq:gamma}) is regularized, and the short-distance cut-offs $\alpha$ and $a$ are sent to zero {\em before} taking the limit of a delta-function tunneling term. 

An expression similar to Eq.\ \eqref{eq:nowick} previously appeared in Refs. \cite{chamon95,safi01}, but the inconsistency with free fermion was not discussed there. An inconsistency between a fermionic calculation and a refermionized calculation after bosonization was recently reported by Shah and Bolech \cite{shah15,bolech15}, also for a problem with an unregularized tunnel Hamiltonian. We suspect that the these discrepancies can also have their origin in the uncontrolled regularization implied by the bosonization procedure.

\subsection{Perturbation theory with regularized tunneling term }\label{app:reg}

We now rederive Eqs. \eqref{eq:I1}, \eqref{eq:I2} and \eqref{eq:I3} up to fourth order in $t$, using standard perturbation theory on the Keldysh contour. This calculation shows that the fermionic and bosonized approaches give identical results if the tunneling term is properly regularized and the ultraviolet cut-offs of the theory are sent to zero before the regularization parameter $\delta$ of the tunneling Hamiltonian.

The two regularization procedures \eqref{eq:ch1} and \eqref{eq:ch2} regularize the current operator in two different ways,
\begin{align}
  I^{({\rm I})} &= \frac{2e}{\hbar} \int dxdx'f(x)f(x')\mbox{Re}\Big[\gamma(t)\Lambda^<(0,x;0,x')\Big], \label{eq:B1} \\
  I^{({\rm II})} &= \frac{2e}{\hbar} \int dxf(x)\mbox{Re}\Big[\gamma(t)\Lambda^<(0,x;0,x)\Big], \label{eq:B2}
\end{align}
where the correlation function $\Lambda$ was defined in Eq.\ (\ref{eq:Lambdadef}) and where the time dependence of the tunneling amplitudes $\gamma$ was introduced as in Eq.\ (\ref{eq:gammat}). We first discuss the calculation of the current to leading order in  perturbation theory, and then discuss the subleading contribution. To avoid spurious repetitions, intermediate results will be given for Choice II of the regularization only, and we discuss Choice I at the end of this Subsection.

\subsubsection{Leading order in tunneling amplitude}\label{app:led}

Choice II in Eq.\ \eqref{eq:ch2} has two variants, depending on the relative chirality of fermions in the contact and in the wire. Following Eq.\ (\ref{eq:contact}) we use the chirality label $r = -$ for case IIa and $r = +$ for case IIb. The leading (second) order contribution to the correlation function $\Lambda$ in Eq. (\ref{eq:B2}) reads
\begin{equation}
\begin{split}
\Lambda^<(0,x;0,x) &=\frac\gamma\hbar\int dx'f(x')\int d\tau e^{-i\Omega \tau}\\
&\Big[C^{\rm T}_r(x-x',-\tau)G^{<}(x'-x,\tau)\\&-C^<_r(x-x',-\tau)G^{\tilde {\rm T}}(x'-x,\tau)\Big],
\end{split}
\end{equation}
where the $r$ dependence is in the Green functions $C_r$ of the contact only. The superscripts $\rm T$ and $\tilde {\rm T}$ refer to time ordered and anti-time ordered correlation functions, respectively, see also the discussion following Eq.\ (\ref{eq:Lambdadef}). Inserting this expression into Eq. \eqref{eq:B2} and transforming to frequency representation gives
\begin{equation}\label{eq:ilofrec}
\begin{aligned}
I^{(2)}=\frac{2e}{\hbar^2}&\gamma^2\mbox{Re}\int dxdx' f(x)f(x')\int \frac{d\omega_1}{2\pi}\\
&\Big[C^{\rm T}_r(x-x',\omega_1)G^{<}(x'-x,\omega_1-\Omega)\\
&-C^<_r(x-x',\omega_1)G^{\tilde {\rm T}}(x'-x,\omega_1-\Omega)\Big]\,.
\end{aligned}
\end{equation}
The limit $\alpha\rightarrow0 $ has to be taken first and can be carried out using the explicit expressions for the correlation functions given in Eq.\ \eqref{eq:greenfree}. For example,
\begin{align}\label{eq:a0}
C^>_r(x,t)&=-\frac1{2\pi v}\mathcal P\left[\frac1{t-r x/v}\right]-\frac i {2v}\delta(t-r x/v)\,,
\end{align}
in which the symbol $\mathcal P$ stands for the Cauchy principal part. 
The relevant Fourier transforms of the Green functions in Eq.\ (\ref{eq:ilofrec}) are
\begin{align}\label{eq:greenfrec}
C^>_r(x,\omega)&=-\frac iv\theta(\omega)e^{e\omega r x/v}\,, \nonumber \\
C^<_r(x,\omega)&=\frac iv\theta(-\omega)e^{e\omega r x/v}\,, \nonumber \\
C^{\rm T}_r(x,\omega)&=\frac 12\Big[C^>_r(x,\omega)+C^<_r(x,\omega)\Big]+A_{v,r}(x,\omega)\,, \nonumber \\
C^{\tilde {\rm T}}_r(x,\omega)&=\frac 12\Big[C^>_r(x,\omega)+C^<_r(x,\omega)\Big]-A_{v,r}(x,\omega)\,, \nonumber \\
A_{v,r}(x,\omega)&=-\frac i{2v}\mbox{sgn}(r x)e^{i\omega r x/v},
\end{align}
with similar expressions for the wire Green functions $G$.
We specialize to the case $\Omega>0$, so that Eq.\ \eqref{eq:ilofrec} simplifies to 
\begin{equation}\label{eq:ilofrec2}
\begin{aligned}
I^{(2)}=&\, \frac{e}{\hbar^2}\gamma^2\mbox{Re}\int d xdx'  f(x)f(x')\int\frac{d\omega_1}{2\pi}
\\&\left[C^>_r(x-x',\omega_1)G^<(x'-x,\omega_1-\Omega) \right. \\
& \mbox{} +	2 A_{v,r}(x-x',\omega_1)G^<(x'-x,\omega_1-\Omega) \\
&\left. \mbox{} + 2 C^<_r(x-x',\omega_1)A_{u,+}(x'-x,\omega_1-\Omega)\right]\,.
\end{aligned}
\end{equation}
One first notices that all the $x$ occurring in the exponential factors, see Eq.\ \eqref{eq:greenfrec}, can be sent to zero as they are all of order $\delta$. Then only the functions $A$ depend on position (but not frequency). Being antisymmetric functions of $x$ (after $x$ has been sent to zero in the exponent), they, however, always integrate to zero to this order for any regularization of the tunneling term for which the function $f(x)$ is an even function of its argument.
A similar reasoning can be applied to Choice I in Eq. \eqref{eq:ch1}. In both cases the result for the current Eq. \eqref{eq:ilot} is then recovered to second order in $t$.

\subsubsection{Next-to-leading order}

For regularization Choice II, the next-to-leading order contribution reads
\begin{widetext}
\begin{equation}
\begin{aligned}\label{eq:4ch2}
I^{(4)} =& \frac{e\gamma^4}{\pi\hbar^4}\mbox{Re}\int d xdx_1dx_2dx_3 f(x)f(x_1)f(x_2)f(x_3)\sum_{\eta_1\eta_2\eta_3}\eta_1\eta_2\eta_3 \int d\omega_1  \\
&\mbox{} \times C^{+\eta_1}_r(x-x_1,\omega_1)G^{\eta_1\eta_2}(x_1-x_2,\omega_1-\Omega)C^{\eta_2\eta_3}_r(x_2-x_3,\omega_1)G^{\eta_3-}(x_3-x,\omega_1-\Omega)\,.
\end{aligned}
\end{equation}
The sum over the Keldysh labels $\eta = \pm$ leads to eight terms which can be all expressed in terms of the functions $C^{>/<}$, $G^{>/<}$, and $A$. In the previous Section we showed that for both regularizations terms linear in $A$ always integrate to zero, and the same applies for terms involving three factors $A$. After some algebra one arrives at
\begin{align}\label{eq:i4inter}
I^{(4)}=& \, \mbox{} \frac{e\gamma^4}{\pi\hbar^4}
  \int d xdx_1dx_2dx_3 f(x)f(x_1)f(x_2)f(x_3) \int d\omega_1 \nonumber \\ & \mbox{} \times
  \Big[-\frac14C^>_r(x-x_1,\omega_1)G^<(x_1-x_2,\omega_1-\Omega)C^>_r(x_2-x_3,\omega_1)G^<(x_3-x,\omega_1-\Omega)
\nonumber \\ & \mbox{} +A_{v,r}(x-x_1)A_{u,+}(x_1-x_2) C^>_r(x_3-x_2,\omega_1)G^<(x_3-x,\omega_1-\Omega)\Big].
\end{align}
For the first term in the integral the limit $\delta\rightarrow 0 $ can be taken before the frequency integration, and one finds the contribution $-8  e^2 t^4 V/h$ to the current, which is the correction one finds in Choice I, see Eqs. \eqref{eq:I1} and \eqref{eq:i4fermion}. For the second term we need to consider the explicit form of the regularizing functions $f(x)$. For Choice II the spatial dependence of $C^>_r$ and $G^<$ can be neglected in the limit $\delta\rightarrow0$ and the result depends on the relative chirality of the two wires only,
\begin{equation}
\begin{aligned}\label{eq:I4II}
\frac {I^{(4)}}{V}&=-\frac{8t^4e^2}{h}-\frac{8t^4e^2}{h}\frac r{(2\delta)^3}\int_{-\delta}^\delta d x_1 dx_2 dx_3\,\mbox{sgn}(x_1-x_2)\,\mbox{sgn}(x_2-x_3)\\
&=-\frac{8t^4e^2}{h}\left(1-\frac r3\right),
\end{aligned}
\end{equation}
\end{widetext}
recovering the fourth-order-in-$t$ contributions to the corrections \eqref{eq:I2} and \eqref{eq:I3} for $r=-$ and $r=+$, respectively.
Repeating the same procedure for Choice I in Eqs. \eqref{eq:ch1} and \eqref{eq:B1}, one finds that the second term in Eq. \eqref{eq:i4inter} is always zero, in agreement with the fourth order expansion of the current in Eqs. \eqref{eq:I1} and \eqref{eq:i4fermion}, independently of the respective chirality of electrons. This concludes our derivation of the regularization-dependent currents given in Eqs. \eqref{eq:conductances} with regularized perturbation theory.

\section{Consistency  with functional bosonization}\label{sec:funcbos}

Although regularization of the tunneling term resolves the discrepancies between the fermionic and bosonic approaches, this resolution is not very satisfying, since taking account of the regularization of the tunneling term comes at the cost of significant technical complications. Whereas every choice of a bosonization prescription inevitably comes with a short-distance cut-off and, hence, with an implicit regularization of the tunneling Hamiltonian (\ref{eq:gamma}), there would be less of a problem if this implicit regularization is the same for fermionic and bosonic approaches. In this section we show that the functional bosonization prescription, a technique first devised  by Yurkevich and Lerner \cite{yurkevich02,lerner05,grishin04} and extended to out of equilibrium situations by Gutman, Gefen, and Mirlin in Ref.\ \cite{gutman10}, satisfies this property.

Starting point of the functional bosonization procedure is the introduction of auxiliary bosonic fields through a Hubbard-Stratonovich transformation, followed by a gauge transformation of the fermionic fields, which is chosen in such a way that the transformed fermionic fields have no residual interactions. To keep the discussion general, in this Section we will consider a wire with right moving and left moving electrons, which are labeled by the index $r = \pm 1$ respectively. The auxiliary bosonic field is denoted $\phi_r$, and the gauge transformation of the fermionic fields is of the form
\begin{align}\label{eq:funcbos}
\psi_r\rightarrow\psi_r\,e^{i\theta_r}\,,
\end{align}
in which  the bosonic fields $\theta_r$ is chosen in such a way, that the gauge-transformed fermions are non-interacting. 

A comparison of Eqs.\ \eqref{eq:bos} and \eqref{eq:funcbos} reveals the main differences with standard bosonization: Functional bosonization  maps  interacting fermions onto \textit{free} fermions and relegates all interaction effects to the bosonic fields $\theta_r$. Functional bosonization then has the double advantage to avoid the use of Klein factors and make a clear separation between fermionic and bosonic degrees of freedom. As a consequence, one finds that interactions affect free fermionic correlation functions by \textit{global prefactors}, even at a finite value of the short-distance cut-offs, see Eqs. \eqref{eq:gintfbos}, \eqref{eq:4pfunc} and \eqref{eq:4pm}. This is different from Eq. \eqref{eq:4p}, derived within standard bosonization. In particular, the functional bosonization prescription straightforwardly reproduces Wick's theorem in the limit of non-interacting fermions.

An additional feature of the functional bosonization approach is that it naturally allows to distinguish interaction and band-width cutoffs. The calculation presented in this Section revisits well known calculations  \cite{yurkevich02,lerner05}, extending these by the disambiguation between interaction and band-width cutoffs in deriving $N$-point correlation functions within this formalism. We here briefly sketch the main steps of the derivation. All additional information about calculations are provided in Appendix \ref{app:funcbos}.

Functional bosonization is operated in field theory language. The Keldysh action of a general interacting wire with linearized spectrum reads
\begin{equation}
\begin{split}
\mathcal S =&\, \mathcal S_0+\mathcal S_1\,,\\
\mathcal S_0 =& \int dxdxdtdt'
  \\ & \mbox{} \times \sum_r \overline\psi_r (x,t) G_r^{-1}(x-x',t-t') \psi_r(x',t')\,,\\
\mathcal S_1 =& -\frac12\int dxdx'dtdt' \\ & \mbox{} \times \sum_{rr'} 
  n_r(x,t)V_{rr'}(x-x',t-t') n_{r'}(x',t'),
\end{split}
\end{equation}
in which all time integrals are performed on the Keldysh contour. The index $r=\pm 1$, for left and right moving electrons, respectively. The electron density is $n_r=\overline \psi_r\psi_r$. We also assume implicitly the standard Keldysh matrix structure in which $\psi=(\psi^+,\psi^-)$ are vectors of fermionic Grasmann variables defined on the upper and lower Keldysh branches. The $2 \times 2$ matrix Green function 
\begin{align}
G^{\CKeldysh}_r&=\left(\begin{array}{cc}G^{\rm T}_r&G^<_r\\G^>_r&G^{\tilde {\rm T}}_r\end{array}\right)=-i\av{ \mathcal T_{\CKeldysh} \psi_r\overline\psi_r}
\end{align}
collects all free fermion Green functions defined in Eq.\ \eqref{eq:greenfree}. For the interaction matrix, we adopt the conventional ``g-ology'' labeling \cite{emery79,solyom79}
\begin{equation}\label{eq:vmatrix}
V_{rr'}(x-x',t-t')=\delta (t-t')\left[\begin{array}{cc}
g_4(x-x')&g_2(x-x')\\g_2(x-x')&g_4(x-x')
\end{array}\right]\,,
\end{equation}
in which $g_{4,2}$ describe forward- and backscattering interaction between electrons  respectively. The first step consists in decoupling fermion densities in $\mathcal S_1$ by introducing  auxiliary fields $\phi_r$ via the Hubbard-Stratonovich transformation
\begin{align}
e^{-\frac i{2}\int n_\eta V_{\eta\eta'}n_{\eta'}}=\int \mathcal D[\phi]e^{\frac i{2}\int(\phi_\eta V_{\eta\eta'}^{-1}\phi_{\eta'}-\phi_\eta n_\eta)}\,.
\end{align}
The action then takes the form 
\begin{equation}
  \label{eq:actionS}
\begin{aligned}
\mathcal S[\psi,\overline\psi,\phi]=\frac12&\sum_{\eta\eta'}\int\phi_r V_{rr'}^{-1}\phi_{r'}+\\&+\sum_\eta\int\,\overline \psi_r(i\partial_t+ir u\partial_x-\phi_r)\psi_r\,,
\end{aligned}
\end{equation}
in which we made explicit the formal identity $G^{-1}_r=(i\partial_t+iru\partial_x)$. This identity has to be understood with the correct Keldysh structure \cite{kamenev11}. The coupling between boson and fermion fields can be gauged out with Eq. \eqref{eq:funcbos} under the condition 
\begin{equation}\label{eq:cond}
\begin{aligned}
D_r\theta_r ^\pm&=-\phi_r^\pm\,,&D_r^{-1}&=\big[\partial_t+r u\partial_x\big]\,.
\end{aligned}
\end{equation}

The Jacobian of the gauge transformation accounts for an additional contribution to the action, 
\begin{align}\label{eq:trasfj}
\begin{split}
\int \mathcal D [&\overline \psi,\psi]e^{\frac i\sum_r\int\overline \psi_r(i\partial_t+ir u\partial_x-\phi_n)\psi_r} \\&\longrightarrow\int \mathcal D [\overline \psi,\psi]e^{ i\sum_r\int\overline \psi_r(i\partial_t+iru\partial_x)\psi_r } \mathcal J[\phi]\,,
\end{split}
\end{align}
with 
\begin{align}\label{eq:jac}
\ln\mathcal J[\phi]=\sum_r\mbox{Tr}\ln[1-G_r \phi_r] \,.
\end{align}
Equation \eqref{eq:trasfj} describes the decoupling of free fermion fields with linear dispersion and boson fields accounting for interactions. Dzyaloshinki and Larkin showed that the Random Phase Approximation (RPA) for the bosonic field is exact  \cite{dzyaloshinskii74}. As a consequence, the determinant contribution to the action \eqref{eq:jac} leads to quadratic contributions in the boson fields $\phi_r$ only. Here it is practical to perform the Keldysh rotation to ``classical'' and ``quantum'' boson fields
\begin{align}
\left(\begin{array}{c}
\phi^c_r\\\phi^q_r
\end{array}\right)
&=\frac1{\sqrt2}\left(\begin{array}{cc}
1&1\\1&-1
\end{array}\right)
\left(\begin{array}{c}
\phi^+_r\\\phi^-_r
\end{array}\right)\,,
\end{align}
which we will adopt for the remainder of this Section. We also introduce $\sigma_0=\mathbbm 1$ and $\sigma_1=\sigma_x$. The Jacobian (\ref{eq:jac}) simplifies to
\begin{equation}
\ln \mathcal J[\phi_r]=-\frac14\mbox{Tr}\left[G_r\phi_r^\alpha\sigma_\alpha G_r\phi_r^\beta\sigma_\beta\right], \label{eq:lnJ}
\end{equation} 
where we assumed the implicit summation of repeated indices $\alpha$ and $\beta$.

An effective action for the bosonic fields $\phi_r$ can now be derived easily. The result takes the most compact form if we define the polarization functions $\Pi$ as $2 i$ times the coefficients of products of classical and quantum fields in Eq.\ (\ref{eq:lnJ}). Carrying out the trace in Eq.\ (\ref{eq:lnJ}) in reciprocal space, one then finds 
\begin{equation}\label{eq:pol}
\begin{split}
\Pi^{\rm A}_r&=-\frac i2\big[G^{\rm K}G^{\rm R}+G^{\rm A}G^{\rm K}\big] \\ & =\frac r{2\pi}\frac p{\omega-i0^+- r u p}\,,\\
\Pi^{\rm R}_r&=-\frac i2\big[G^{\rm K}G^{\rm A}+G^{\rm R}G^{\rm K}\big] \\ & =\frac r{2\pi}\frac p{\omega+i0^+- r u p}\,,\\
\Pi^{\rm K}_r&=-\frac i2\big[G^{\rm K}G^{\rm K}+G^{\rm A}G^{\rm R}+G^{\rm R}G^{\rm A}\big]\\
&=\coth\left(\frac\omega{2T}\right)\Big[\Pi^{\rm R}_r(p,\omega)-\Pi^{\rm A}_r(p,\omega)\Big],
\end{split}
\end{equation}
where the superscripts ${\rm A}$, ${\rm R}$, and ${\rm K}$ refer to the advanced, retarded, and Keldysh components \cite{kamenev11}.
Substituting this into the action (\ref{eq:actionS}), the effective action of the $\phi$ fields can be expressed as  
\begin{equation}
\mathcal S[\phi] =\phi_r\mathcal V_{rr'}^{-1}\phi_{r'},
\end{equation}
with
\begin{equation}\label{eq:vrrinv}
\mathcal V_{rr'}^{-1} =\left[\begin{array}{cc}
0&V^{-1}_{rr'}-\delta_{rr'}\Pi^{\rm A}\\
V^{-1}_{rr'}-\delta_{rr'}\Pi^{\rm R}&-\delta_{rr'}\Pi^{\rm K}
\end{array}\right].
\end{equation}
The inverse matrix $\mathcal V$ is the correlation function of the fields $\phi$ and it is given explicitly in reciprocal space in Appendix \ref{app:funcbos}. It is directly related to the correlation matrix imposed on the fields $\theta$ appearing in Eqs. \eqref{eq:funcbos}  and \eqref{eq:cond}, through \cite{schneider11} 
\begin{widetext}
\begin{equation}\label{eq:link}
\left(\begin{array}{c}\theta_c (x,t) \\ \theta_q(x,t) \end{array}\right)=-\int dx' dt' \left(\begin{array}{cc}D^{\rm R}(x-x',t-t') &D^{\rm K}(x-x',t-t')\\0&D^{\rm A}(x-x',t-t')\end{array}\right) \left(\begin{array}{c}\phi_c(x',t')\\\phi_q(x',t')\end{array}\right).
\end{equation}
The choice of signs in the matrix is such to fulfill the correct boson causality condition for the $\theta$ fields, see Appendix \ref{app:funcbos}. After quite tedious, but standard, algebra --- details in Appendix \ref{app:funcbos} ---, the correlation functions 
\begin{equation}\label{eq:greentheta}
F_{rr'}^{\CKeldysh} =-i\av{ \mathcal T_{\CKeldysh} \theta_r\theta_{r'}}
\end{equation}
are derived in real space. For example, the ``greater'' correlation functions read 
\begin{align}\label{eq:fm1}
&F^>_{rr}(x,t)=-i\int dp\frac {e^{ipx}}p\left\{e^{-iwpt}\big[1+n_B(wp)\big]\frac{(1+r K)^2}{4K}+e^{iwpt}n_B(wp)\frac{(1-r K)^2}{4K}-r e^{-ir upt}\big[1+n_B(r up)\big]\right\}\,,\\
\label{eq:fm2} 
&F^>_{r,-r}(x,t)= -i\int dp\frac {e^{ipx}}p\frac{1-K^2}{4K}\left\{e^{-iwpt}\big[1+n_B(wp)\big]+e^{iwpt}n_B(wp)\right\}\,,
\end{align}
\end{widetext}
in which $n_B(\omega)=(e^{\beta\omega}-1)^{-1}$ is the Bose-Einstein distribution function,
\begin{equation}
\label{eq:w}w(p) =u\sqrt{\left[1+\frac{g_4(p)}{2\pi u}\right]^2-\left[\frac{g_2(p)}{2\pi u}\right]^2}
\end{equation}
is the velocity of the collective modes induced by interactions, 
and
\begin{equation}
\label{eq:K}K(p) =\sqrt{\frac{2\pi u+g_4(p)-g_2(p)}{2\pi u+g_4(p)+g_2(p)}}
\end{equation}
is the Luttinger parameter. In the absence of interactions one has $K=1$, whereas generally $K<1$ ($K>1$) for repulsive (attractive) interactions \cite{giamarchi04}. 

The momentum dependence of $g_{2}$ and $g_{4}$ allows for the natural introduction of separate cut-offs for interaction and band width. For any finite-range  interaction, $g_{2,4}(p)\neq0$ only on a finite support, whereas $g_{2,4}(p) \to 0$ in the limit $|p|\rightarrow \infty$. As a consequence,  $w\rightarrow u$ and $K\rightarrow 1$ for  $|p|\rightarrow \infty$. This ensures the convergence of both integrals in Eqs. \eqref{eq:fm1} and \eqref{eq:fm2}. [Note that often the functions $g_2$ and $g_4$ are considered as momentum independent, and a single ultraviolet cutoff $a$ is introduced to ensure converge of both (free) fermion and boson correlation functions. That there is actually no reason to make this assumption was pointed out in some of the early works on bosonization, focusing on two-point correlation functions  \cite{dzyaloshinskii74,voit95,schonhammer97}.] We thus assume momentum-independent interaction parameters in Eqs.\ \eqref{eq:w} and \eqref{eq:K} and introduce an exponential cutoff $e^{-\varepsilon|p|}$ in the integrals Eqs. \eqref{eq:fm1} and \eqref{eq:fm2}, where the ultraviolet cut-off $\varepsilon$ is different from the ultraviolet cut-off $\alpha$ used for the free-fermion correlation functions. In fact, the ultraviolet cut-off $\alpha$ can be sent to zero at the end of the calculation \cite{vondelft98}, whereas the cut-off $\varepsilon$ for the electron-electron interactions should remain finite throughout the calculation. An example pointing out the different roles of the two cut-offs can be found in Sec.\ \ref{sec:int}.

Performing the momentum integrations with the ultraviolet cut-off as described above, we find that the greater Green functions of the boson fields become
\begin{widetext}
\begin{align}
F^>_{rr}(x,t)&=-i\left\{\frac{(1+\eta K)^2}{4K}\ln\frac\varepsilon{\varepsilon-i(x-wt)}+\frac{(1-r K)^2}{4K}\ln\frac\varepsilon{\varepsilon+i(x+wt)}-\ln\frac\varepsilon{\varepsilon-i(r x-u t)}\right\}\,,\\
F^>_{-rr}(x,t)&=-i\frac{1-K^2}{4K}\left\{\ln\frac\varepsilon{\varepsilon-i(x-wt)}+\ln\frac\varepsilon{\varepsilon+i(x+wt)}\right\}\,.
\end{align}
\end{widetext}
The lesser functions are readily extracted by using $F^<_{rr'}(x-x',t-t')=-[F_{rr'}^>(x-x',t-t')]^*$. By applying standard properties of the averages of exponentials of fields with quadratic actions, the full fermion Green functions on the Keldysh contour are now easily derived \cite{dzyaloshinskii74,voit95,schonhammer97},
\begin{align}\label{eq:gintfbos}
G^{\CKeldysh}_r(x-x',t-t') \mbox{} = &\, \mbox{} \frac1{2\pi}\frac{\varepsilon-is_{\CKeldysh}(t)(r x-u t)}{\varepsilon(r x-ut+i\alpha s_{\CKeldysh}(t))} \nonumber \\ & \mbox{} \times
  \prod_{\pm} \left(\frac\varepsilon{\varepsilon \mp is_{\CKeldysh}(t)(x \mp wt)}\right)^{\frac{(1 \pm r K)^2}{4K}},
\end{align}
where we point to the appearance of separate cut-offs for the fermionic (band width) and bosonic (interaction) degrees of freedom. A difference with the textbook presentations of the bosonization procedure is that in Eq.\ (\ref{eq:gintfbos}) the sound velocity of the collective modes $w$ does {\em not} replace the Fermi velocity $u$ of the original fermions \cite{giamarchi04,vondelft98,fradkin91}, although the velocity $u$ drops out of the expression for distances larger than  $\varepsilon$.

Notice that the sole effect of interactions is to multiply the free fermion correlation functions by a prefactor. This is a general feature, which  also applies to the four-point and higher order correlation functions. Explicitly, for the four-point correlation functions with all fields evaluated at the same spatial position one finds 
\begin{widetext}
\begin{equation}\label{eq:4pfunc}
\begin{split}
\av{- {\mathcal T}_{\CKeldysh} \psi_r(1)\psi_r(2)\overline\psi_r(3)\overline\psi_r(4)}=&\, \Big[G^{\CKeldysh}_{0r}(1-4)G^{\CKeldysh}_{0r}(2-3) - G^{\CKeldysh}_{0r}(1-3)G^{\CKeldysh}_{0r}(2-4)\Big]\times\\&\times\frac{f_{w}^{\CKeldysh}(t_1-t_3)f_{w}^{\CKeldysh}(t_1-t_4)f_{w}^{\CKeldysh}(t_2-t_3)f_{w}^{\CKeldysh}(t_2-t_4)}{f_{w}^{\CKeldysh}(t_1-t_2)f_{w}^{\CKeldysh}(t_3-t_4)}\,,
\end{split}
\end{equation}
\end{widetext}
where 
\begin{align}
f_{w}^{\CKeldysh}(t)=&\, \mbox{} e^{iF^{\CKeldysh}(0,t)} \nonumber \\=&\, \mbox{} 
  \frac{\varepsilon+iuts_{\CKeldysh}(t)}{\varepsilon+iwts_{\CKeldysh}(t)}\left(\frac\varepsilon{\varepsilon+iwts_{\CKeldysh}(t)}\right)^{\frac{(1-K)^2}{2K}}
\end{align}
and $G^{\CKeldysh}_{0r}$ is the contour-ordered Green function for free fermions. As advertised, Eq.\ \eqref{eq:4pfunc} is the product of a free fermion part, for which Wick's theorem applies, and a global prefactor, which accounts for all interaction effects. The global prefactor simplifies to a factor one for $K=1$, so that Wick's theorem is automatically satisfied in the limit of no electron-electron interactions. Given the nature of the gauge transformation Eq. \eqref{eq:funcbos}, the same applies to all $N$-point correlation functions when calculated in the functional bosonization formalism. It follows that in the non-interacting limit $K \to 1$ physical observables, such as the conductance \eqref{eq:I1}, are the same in the functional bosonization formalism as in the fermionic approach. 

For completeness, we also report the correlation function between counter-propagating fermions, again evaluated at equal positions,
\begin{widetext}
\begin{equation}\label{eq:4pm}
\begin{split}
\av{- {\mathcal T}_{\CKeldysh} \psi_r(1)\psi_{-r}(2)\overline\psi_{-r}(3)\overline\psi_r(4)}&= G_{0r}^{\CKeldysh}(1-4)G_{0-r}^{\CKeldysh}(2-3)\frac{g_{w}^{\CKeldysh}(t_1-t_3)f_{w}^{\CKeldysh}(t_1-t_4)f_{w}^{\CKeldysh}(t_2-t_3)g_{w}^{\CKeldysh}(t_2-t_4)}{g_{w}^{\CKeldysh}(t_1-t_2)g_{w}^{\CKeldysh}(t_3-t_4)}\,.
\end{split}
\end{equation}
\end{widetext}
Here the function $g_{w}^{\CKeldysh}(t)$ is defined as
\begin{equation}
g_{w}^{\CKeldysh}(t)=\left(\frac{\varepsilon}{\varepsilon+iwts_{\CKeldysh}(t)}\right)^{\frac{1}{2K}(K^2+1)(1-K)}\,. 
\end{equation}
Notice that  the long distance behavior of this correlation function has a power law decay with a different exponent from the one present in the co-propagating correlation functions Eqs. \eqref{eq:gintfbos} and \eqref{eq:4pfunc}. 



\section{Application: tunneling current from functional bosonization}\label{sec:int}

As a simple application of the functional bosonization approach, we reexamine the calculation of the tunneling current in the presence of interactions in the wire to leading order in the tunneling amplitude $t$. Our aim is to show how the ultraviolet cut-offs $\alpha$ for the band width and $\varepsilon$ for the interaction range appear in the tunnel current calculations, and to clarify which of these is the cut-off that appears in the known result from the literature Eq. \eqref{eq:2}. As before, we will not regularize the tunneling Hamiltonian in this calculation, and instead rely on the regularization from the ultraviolet cut-offs $\alpha$ and $\varepsilon$.

Without regularization of the tunneling Hamiltonian, all fields are evaluated at the same location. The Green functions \eqref{eq:gintfbos} fulfill $G^>(0,t)^*=G^<(0,t)$ and $G^>(0,-t)=-G^<(0,t)$. After some manipulations and taking explicitly the real part in Eq. \eqref{eq:ilo}, one can recast the expression for the current as an integral in frequency space of exclusively greater and lesser Green functions 
\begin{equation}\label{eq:ilofin}
\begin{aligned}
  I^{(2)} =\frac{2 \gamma^2}{2\pi \hbar^2 }\int d \omega_1\Big[&C^>(\omega_1)G^<(\omega_1-\Omega)-\\&-C^<(\omega_1)G^>(\omega_1-\Omega)\Big]\,.
\end{aligned}
\end{equation}
For small values of the interaction cut-off $\varepsilon$ the Fourier transform of the greater Green function Eq. \eqref{eq:gintfbos} reads (see Appendix \ref{app:funcbos} for details)
\begin{align}\label{eq:gfrec}
G^>(x=0,\omega)&=-i\frac{\varepsilon^{\nu-1}}{w^\nu\Gamma(\nu)}\omega^{\nu-1}\theta(\omega)\,, 
\end{align}
where $G^<(\omega)=G^>(-\omega)^*$, $\Gamma$ is the Gamma function, and
\begin{align}
\nu=\frac12\left(K+\frac1K\right).
\end{align}
In this expression the limit $\alpha \to 0$ has already been taken. The dependence on the Fermi velocity $u$ for the free fermions, still present in Eq. \eqref{eq:gintfbos}, has disappeared, as a consequence of considering all fields at the same point. As discussed above, the regularization of the tunneling Hamiltonian has no consequence to leading order in perturbation theory, but it does for the higher ones. 

For a positive bias $\Omega>0$ only the first term in Eq. \eqref{eq:ilofin} has a nonzero contribution and one finds
\begin{align}\label{eq:ilointf}
  I^{(2)} = \frac{4 t^2 V e^2}{h} \frac1{\Gamma(\nu+1)}\frac{u}{w}\left(\frac{\varepsilon eV}{\hbar w}\right)^{\nu-1}\,,
\end{align}
with $t$ defined in Eq.\ (\ref{eq:tdef}). Essentially this is the same result as in Eq.\ (\ref{eq:2}), with an explicit evaluation of the energy scale $\Lambda = \hbar w/\varepsilon$. The same result is found if the ultraviolet cut-offs $\alpha$ and $\varepsilon$ are taken to be equal. The present calculation underlines the different roles of the two cut-offs and shows that the ultraviolet cut-off $\alpha$ for the band width does not appear in the final result for the tunneling current.


\section{Conclusions}

In this paper we considered the tunneling current between a metallic point-like contact and a one-dimensional wire to higher order in perturbation theory. We re-examined the standard tunneling Hamiltonian  Eq. \eqref{eq:gamma}, commonly considered as a perturbation  in this kind of problems. In the absence of interactions, scattering theory allows us to find the current to arbitrary order in the tunneling amplitude $\gamma$. We showed that the tunneling current is strongly sensitive to the regularization scheme in the strong tunneling limit and pointed out that discrepancies between regularization schemes appear only beyond the leading order expansion in the tunneling amplitude $\gamma$ of the current.

The need to regularize the tunneling term is no longer apparent once an ultraviolet cutoff $\alpha$ is inserted in fermion correlation functions. The same applies when the electrons are described by boson fields, using the bosonization formalism. For free fermions, we showed that this procedure corresponds to a specific regularization of the tunneling Hamiltonian, but it is not consistent with the standard  regularization choice in the bosonization formalism. The reason is that the utilization of an ultraviolet cutoff $a$ for bosonic fields is responsible for a violation of Wick's theorem in the short-time limit. This leads to different results for the tunneling current to higher orders in perturbation theory even in the non-interacting limit. Our results suggest that this regularization inconsistency may also be responsible for other discrepancies recently pointed out in the literature \cite{shah15,bolech15}. On the other hand, with an explicit calculation we showed that the regularization of the tunneling term fully lifts any inconsistencies between free fermions and standard bosonization, as it allows to avoid any uncontrolled ultraviolet regularizations of the tunneling term in both cases. 

The regularization of the tunneling term leads to quite involved calculations and it is important to rely on  bosonization prescriptions which do not lead to uncontrolled modification of the underlying fermionic model. We showed how functional bosonization allows to resolve the inconsistency between free and bosonized fermions and also allows to distinguish, in an intuitive way, between the ultraviolet regularization necessary to account for the approximation of a linear dispersion relation and the short-distance scale imposed by interactions.

We hope that our work will contribute to a consistent theory of higher-order tunneling processes for interacting electrons.


\section{Acknowledgments}
We thank M. Schneider for bringing our attention to the functional bosonization technique, D. Bercioux for contributions at the initial stage of the work, and the Alexander von Humboldt foundation for financial support.  We also thank J.\ von Delft, L.\ Glazman, Y.\ Gefen, H.\ Saleur and F.\ von Oppen for valuable discussions. 


\appendix

\section{Tunneling current from direct solution of the Schr\"odinger equation}\label{app:scatt}

The conductances in Eq. \eqref{eq:conductances} can be derived with the aid of scattering theory, without the need to introduce ultraviolet cut-offs for the fermionic fields. The scattering states are solutions of the Schr\"odinger equation with a properly regularized tunneling Hamiltonian. We present explicit solutions for the three choices I, IIa, and IIb for the regularization, see Fig.\ \ref{fig:regs} and Eqs.\ (\ref{eq:ch1})--(\ref{eq:ch2}). For the function $f(x)$ in Eqs.\ (\ref{eq:ch1})--(\ref{eq:ch2}) we choose a ``box''-like form,
\begin{equation}
  f(x) = \left\{ \begin{array}{ll} 1/2 \delta & \mbox{for $|x| < \delta$},\\
  0 & \mbox{otherwise}. \end{array} \right.
\end{equation}

{\em Choice I.} For the first choice for the regularization of the tunneling term the Schr\"odinger equation reads, for $-\delta < x < \delta$,
\begin{equation}\label{eq:schrod1}
\begin{aligned}
0&= -i \hbar v\partial_x \Psi_{\rm C}(x)+ \frac\gamma{4 \delta^2}\int_{-\delta}^\delta dx' \Psi_{\rm W}(x'),\\
0&= -i\hbar u\partial_x\Psi_{\rm W}(x)+\frac\gamma{4\delta^2} \int_{-\delta}^\delta dx' \Psi_{\rm C}(x'),
\end{aligned}
\end{equation} 
where $\Psi_{{\rm C},{\rm W}}$ are the wavefunctions of the scattering state in the contact ($\rm C$) and wire ($\rm W$), respectively. (Without loss of generality we have set $r=1$.) For a particle incident from the contact we solve these equations with the boundary condition $\psi_{\rm C}(-\delta) = 1/\sqrt{2 \pi v}$, $\psi_{\rm W}(-\delta) = 0$, corresponding to unit incoming flux in the contact. For $-\delta < x < \delta$ the solution reads
\begin{equation}
\begin{aligned}
  \Psi_{\rm C}(x) & = \frac{1}{\sqrt{2 \pi v}}
  \left[ 1 - \frac{t^2(x+\delta)}{4 \delta (1+t^2)} \right], \\
  \Psi_{\rm W}(x) & = - \frac{i}{\sqrt{2 \pi u}}
  \frac{t (x+\delta)}{\delta(1+t^2)},
\end{aligned}
\end{equation}
where $t = \gamma/2 \hbar \sqrt{u v}$. The transmission amplitude is then given by $\psi_{\rm W}(\delta) \sqrt{2 \pi u}$, so that, by the Landauer formula, we find the tunneling conductance
\begin{equation}
  G^{({\rm I})} = \frac{4 t^2 e^2}{h(1 + t^2)^2},
\end{equation}
see Eq.\ (\ref{eq:I1}).

{\em Choice IIa.} In this case, the Sch\"odinger equation reads, for $-\delta < x < \delta$ 
\begin{equation}\label{eq:schrod2a}
\begin{aligned}
0&= i\hbar v\partial_x \Psi_{\rm C}(x)+\frac\gamma{2\delta} \Psi_{\rm W}(x),\\
0&=-i\hbar u\partial_x \Psi_{\rm W}(x)+\frac\gamma{2\delta} \Psi_{\rm C}(x)
\end{aligned}
\end{equation} 
Since the electrons in the contact now propagate in the negative $x$ direction, the boundary condition corresponding to an electron incident from the normal contact reads $\psi_{\rm C}(\delta) = 1/\sqrt{2 \pi v}$, $\psi_{\rm W}(-\delta) = 0$. The solution of the Schr\"odinger equation (\ref{eq:schrod2a}) with this boundary condition is
\begin{equation}
\begin{aligned}
  \Psi_{\rm C}(x) & = \frac{\cosh[t(x/\delta+1)]}{\sqrt{2 \pi v} \cosh 2t}, \\
  \Psi_{\rm W}(x) & = \frac{\sinh[t(x/\delta+1)]}{i \sqrt{2 \pi u} \cosh 2t }.
\end{aligned}
\end{equation}
As before, the transmission amplitude is $\psi_{\rm W}(\delta) \sqrt{2 \pi u}$, so that we find the conductance (\ref{eq:I2}).

{\em Choice IIb.} For this choice of the regularization, the Schr\"odinger equation reads
\begin{equation}\label{eq:schrod2b}
\begin{aligned}
0&=-i\hbar v\partial_x \Psi_{\rm C}(x)+\frac\gamma{2\delta} \Psi_{\rm W}(x),\\
0&=-i\hbar u\partial_x \Psi_{\rm W}(x)+\frac\gamma{2\delta} \Psi_{\rm C}(x),
\end{aligned}
\end{equation} 
and the boundary condition is the same as for Choice I. The solution is
\begin{equation}
\begin{aligned}
  \Psi_{\rm C}(x) & = \frac{\cos[t(x/\delta+1)]}{\sqrt{2 \pi v}}, \\
  \Psi_{\rm W}(x) & = \frac{\sin[t(x/\delta+1)]}{i \sqrt{2 \pi u}}.
\end{aligned}
\end{equation}
Again the transmission amplitude is $\psi_{\rm W}(\delta) \sqrt{2 \pi u}$, which gives the conductance (\ref{eq:I3}).


\section{Intermediate results for the functional bosonization procedure}\label{app:funcbos}

In this appendix we provide details on the derivation of the bosonic correlation functions Eqs. \eqref{eq:fm1} and \eqref{eq:fm2}. 

{\em Derivation of the correlation functions (\ref{eq:vrrinv}).---}
We start by deriving explicit expressions for the correlation functions 
\begin{equation}
  \mathcal V_{rr'} = -i\av{\phi_r\phi_{r'}}.
\end{equation}
The matrix $\mathcal V_{rr'}$ is the inverse of Eq. \eqref{eq:vrrinv} and reads 
\begin{widetext}
\begin{equation}\label{eq:v}
\mathcal V_{rr'}=\left[\begin{array}{cc}
  \sum_{r''} 
  \big(V^{-1}-\Pi^{\rm R}\big)_{rr''}^{-1}\Pi^{\rm K}_{r''}\big(V^{-1}-\Pi^{\rm A}\big)_{r''r'}^{-1} 
  & \big(V^{-1}-\Pi^{\rm R}\big)_{rr'}^{-1}\\
\big(V^{-1}-\Pi^{\rm A}\big)_{rr'}^{-1}&0
\end{array}\right]\,.
\end{equation} 
We switch to reciprocal space, in which 
\begin{equation}
\big[V^{-1}(p)-\Pi^{{\rm R}/{\rm A}}(\omega,p)\big]_{r,r'}=
  \left[\begin{array}{cc}
\frac{g_4}{g^2_4-g^2_2}-\Pi^{{\rm R}/{\rm A}}_+&-\frac{g_2}{g^2_4-g^2_2}\\
-\frac{g_2}{g^2_4-g^2_2}&\frac{g_4}{g^2_4-g^2_2}-\Pi^{{\rm R}/{\rm A}}_-
\end{array}\right],
\end{equation}
where we have suppressed the dependence of the functions $g_{2,4}$ on the momentum $p$. Inverting this equation and inserting it into Eq. \eqref{eq:v} we find that the off-diagonal elements of $\mathcal V_{rr'}$ are given by
\begin{eqnarray}
\mathcal V^{{\rm R}/{\rm A}}_{rr'} &=&
  \big[V^{-1}(p)-\Pi^{{\rm R}/{\rm A}}(\omega,p)\big]^{-1}_{r,r'}
  \nonumber \\ &=&
  \frac1{\left(\frac{g_4}{g^2_4-g^2_2}-\Pi^{{\rm R}/{\rm A}}_+\right)\left(\frac{g_4}{g^2_4-g^2_2}-\Pi^{{\rm R}/{\rm A}}_-\right)-\frac{g_2^2}{(g_4^2-g_2^2)^2}}
\left[\begin{array}{cc}
\frac{g_4}{g^2_4-g^2_2}-\Pi^{{\rm R}/{\rm A}}_-&\frac{g_2}{g^2_4-g^2_2}\\
\frac{g_2}{g^2_4-g^2_2}&\frac{g_4}{g^2_4-g^2_2}-\Pi^{R/A}_+
\end{array}\right].
\end{eqnarray}
\end{widetext}
After some algebra, the following explicit expressions for the retarded/advanced components of $\mathcal V_{rr'}$ are derived:
\begin{equation}
\begin{split}
\mathcal V^{{\rm R}/{\rm A}}_{+-}(\omega,p) =&\, \mathcal V^{{\rm R}/{\rm A}}_{-+}(\omega,p)
  \nonumber \\ =&\, \mbox{}
g_2(p)\frac{\omega_\pm^2-v_F p^2}{\omega_\pm^2-w^2(p)p^2}\,,\\
\mathcal V^{{\rm R}/{\rm A}}_{++}(\omega,p) =&\,
\mathcal V^{{\rm R}/{\rm A}}_{--}(\omega,p) \nonumber \\ =&\, \mbox{}
\frac{g_4(p)(\omega_\pm-r v_Fp)}{\omega^2_\pm-w^2(p)p^2}
  \nonumber \\ & \mbox{} \times
\left[\omega_\pm+r p\left(v_F+\frac1{2\pi}\frac{g^2_4(p)-g^2_2(p)}{g_4(p)}\right)\right],
\end{split}
\end{equation}
in which $\omega_{\pm} = \omega \pm i0$ and we introduced the renormalized velocity of collective modes $w(p)$, see Eq.\ \eqref{eq:w}. One can verify that the Keldysh component $\mathcal V^{\rm K}$ in Eq. \eqref{eq:v} fulfills the general bosonic fluctuation-dissipation relation
\begin{equation}
\mathcal
 V_{rr'}^{\rm K}(\omega,p)=\coth\left(\frac\omega{2T}\right)\Big[\mathcal V^{\rm R}_{rr'}(\omega,p)-\mathcal V^{\rm A}_{rr'}(\omega,p)\Big],
\end{equation}
which completes the calculation of $\mathcal V_{rr'}$.

{\em Derivation of Eq.\ (\ref{eq:greentheta}).---}
Next, we derive the Green function \eqref{eq:greentheta}, starting from Eq. \eqref{eq:link} of the main text. This Green function has the usual bosonic structure, in which, after switching to retarded/advanced/Keldysh components \cite{kamenev11},
\begin{align}
F_{rr'}=
\left(
\begin{array}{cc}
F^{\rm K}_{rr'}&F^{\rm R}_{rr'}\\
F^{\rm A}_{rr'}&0
\end{array}
\right)\,,
\end{align} 
with 
\begin{equation}\label{eq:ththra}
\begin{split}
F_{rr'}^{{\rm R}/{\rm A}}(\omega,p)&=D_r^{{\rm R}/{\rm A}}(\omega,p)\mathcal V_{r'}^{{\rm R}/{\rm A}}(\omega,p)D_{r'}^{{\rm A}/{\rm R}}(-\omega,-p).
\end{split}
\end{equation}
Substitution of the expressions for $D_r$ and $\mathcal V_r$ from the main text then leads to
\begin{equation}\label{eq:hra}
\begin{split}
F_{+-}^{{\rm R}/{\rm A}}(\omega,p)&=
F_{-+}^{{\rm R}/{\rm A}}(\omega,p) \nonumber \\ &=
\frac{g_2}{\omega_\pm^2-p^2w^2}\,,\\
F^{{\rm R}/{\rm A}}_{++}(\omega,p)&=F^{{\rm R}/{\rm A}}_{--}(\omega,p) \nonumber \\ &=
g_4\frac{\omega_\pm+r p\left(v_F+\frac1{2\pi}\frac{g^2_4-g^2_2}{g_4}\right)}{(\omega_\pm ^2-w^2p^2)(\omega_\pm-r v_Fp)}\,.
\end{split}
\end{equation}
Again, the Keldysh component $F^{\rm K}_{rr'}(\omega,p)$ may be calculated from the fluctuation-dissipation relation,
\begin{equation}
\begin{split}
F^{\rm K}_{rr'}(\omega,p)=\coth\left(\frac{\omega}{2T}\right)\Big[F^{\rm R}_{rr'}(\omega,p)-F^{\rm A}_{rr'}(\omega,p)\Big].
\end{split}
\end{equation}

{\em Derivation of Eqs.\ (\ref{eq:fm1}) and (\ref{eq:fm2}).---} 
To derive Eq.\ \eqref{eq:fm1} we return to real space. Using the definitions of the retarded, advanced, and Keldysh components, as well as the fluctuations-dissipation theorem, one has 
\begin{align}
F^>(\omega,p) &=
  \frac12\big(F^{\rm K}(\omega,p)+F^{\rm R}(\omega,p)-F^{\rm A}(\omega,p)\big) \nonumber \\
  &=\big[1+n_B(\omega)\big] (F^{\rm R}(\omega,p)-F^{\rm A}(\omega,p)),
\end{align}
with $n_B(\omega)=(e^{\beta\omega}-1)^{-1}$ the Bose-Einstein distribution. First Fourier transforming in the time domain, we find 
\begin{equation}
\begin{split}
F^>(p,t)
  &=\frac1{2\pi}\theta(t)\int_{\tilde{\mathcal C}_-} dz e^{-i z t}\big[1+n_B(z)\big]F^{\rm R}(z,p)\\&~~-\frac1{2\pi}\theta(-t)\int_{\tilde{\mathcal C}_+} dz e^{-i z t}\big[1+n_B(z)\big]F^{\rm A}(z,p)\,,
\end{split}
\end{equation} 
in which $\tilde{\mathcal C}_+$ and $\tilde{\mathcal C}_-$ denote the standard complex contours closed in the upper/lower plane respectively. The poles of $n_B(\omega)$ do not contribute, because $F^{\rm R}(i\omega)=F^{\rm A}(i\omega)$ for any finite complex frequency. The two integrals lead to the same result for positive and negative $t$, so that
\begin{equation}
\begin{split}
F^>(p,t)&=\frac1{2\pi}\int_{\tilde{\mathcal C}_-} dz e^{-i z |t|}\big[1+n_B(z)\big]F^{\rm R}(z,p)\,.
\end{split}
\end{equation}
For co-moving fields, $r=r'$, the function $F^{\rm R}_{rr}$ appearing in the integrand has poles at $\omega = \pm  w(p)p$ and at $\omega =r v_Fp$, leading to 
\begin{widetext}
\begin{equation}
\begin{split}
F^>_{\eta\eta}(p,t)=-ig_4 & \left\{ \sum_{\pm} 
  e^{\mp iupt}\big[1+n_B(\pm up)\big]\frac{\eta  p\left(v_F+\frac1{2\pi}\frac{g^2_4-g^2_2}{g_4}\right) \pm up}{2up(up \mp \eta v_Fp)} \right. \\
& \left. \mbox{} + e^{-i\eta v_Fpt}\big[1+n_B(\eta v_Fp)\big]\frac{\eta v_F p+\eta  p\left(v_F+\frac1{2\pi}\frac{g^2_4-g^2_2}{g_4}\right)}{(\eta v_Fp-up)(\eta v_Fp+up)}\right\}.
\end{split}
\end{equation}
\end{widetext}
The Fourier transform of the momentum argument then gives Eq.\ \eqref{eq:fm1}. Equation \eqref{eq:fm2} for counter-moving fields is derived in a similar manner.

 
{\em Fourier transform of interacting green functions.}
We conclude by providing some details on the derivation of the Fourier transform of the Green function \eqref{eq:gintfbos}, leading to Eq. \eqref{eq:gfrec}. For $x=0$, Eq. \eqref{eq:gintfbos} reads 
\begin{align}
G^>(0,t)&= \frac {\varepsilon^{\nu-1}}{2 \pi (-ut+i\alpha)}\frac{\varepsilon+iut}{(\varepsilon+iwt)^\nu}\,.
\end{align}
If $\nu$ is a positive integer the Fourier transform can be carried out with standard complex integration techniques, leading to 
\begin{align}
G^>(0,\omega) =&-\frac{\varepsilon^{\nu-1}i}{u(iw)^\nu}\theta(\omega)\left[\frac{\varepsilon e^{-\omega\alpha/u}}{(\frac{i\alpha}u-\frac{i\varepsilon}w)} \right.\\&\left.+\frac1{(\nu-1)!} \frac{d^{\nu-1}}{dz^{\nu-1}}\frac{e^{i\omega z}(iuz+\varepsilon)}{z-\frac{i\alpha}{u}}\right]_{z=\frac{i\varepsilon}{w}}. \nonumber
\end{align}
At this stage the limit $\alpha\rightarrow 0$ can be performed, which gives 
\begin{equation}
\begin{aligned}
G^>(0,\omega)=& -\frac{\varepsilon^{\nu-1}i}{u(iw)^\nu}\theta(\omega)\left[\frac{iu(iw)^{\nu-1}}{(\nu-1)!}e^{-\omega \varepsilon/w}+\right.\\
& \mbox{} +\frac{(iw)^\nu}{\varepsilon^{\nu-1}}+\frac\varepsilon{(\nu-1)!}\left.\frac{d^{\nu-1}}{dz^{\nu-1}}\frac{e^{i\omega z}}{z}\right]_{z=\frac{i\varepsilon}{w}}\,.
\end{aligned}
\end{equation}
In the limit $\varepsilon\rightarrow0$, the two last terms in the above expression simplify to zero,
\begin{equation}\label{eq:form}
\begin{aligned}
  \lim_{\varepsilon \to 0} & \left[
  \frac{(iw)^\nu}{\varepsilon^{\nu-1}}+\frac\varepsilon{(\nu-1)!}\left.\frac{d^{\nu-1}}{dz^{\nu-1}}\frac{e^{i\omega z}}{z}\right|_{z=\frac{i\varepsilon}{w}}
  \right]\\ & =
  \lim_{\varepsilon \to 0} \left[
  \frac{(iw)^\nu}{\varepsilon^{\nu-1}}+\varepsilon\sum_{k=0}^{\nu-1}\frac{(i\omega)^ke^{-\omega\varepsilon/w}(-1)^{\nu-1-k}}{k!\left(i\varepsilon/w\right)^{\nu-k}} \right] \\ & =
  \lim_{\varepsilon \to 0} \left[
  \frac{(iw)^\nu}{\varepsilon^{\nu-1}}+\varepsilon\sum_{k=0}^{\infty}\frac{(i\omega)^ke^{-\omega\varepsilon/w}(-1)^{\nu-1-k}}{k!\left(i\varepsilon/w\right)^{\nu-k}} \right] \\ & = 
\mbox{} 0.
\end{aligned}
\end{equation}
Equation \eqref{eq:gfrec} is recovered upon continuing $\nu$ to continuous values.

\bibliographystyle{apsrev4-1}
\bibliography{bibnoise}

\end{document}